\documentclass[12pt]{article}

\usepackage{epsf}

\begin{document}

\mbox{}\hfill Fermilab-pub-07-413 \\
\mbox{}\hfill MINOS-doc-3412\\
\mbox{}\hfill August 10, 2007

\begin{center}
       Preliminary Results from MINOS 
       on $\nu_\mu$ Disappearance 
       Based on an \\ Exposure of $2.5\times 10^{20}$ 
       $120\;\mbox{GeV}$ Protons on the NuMI Target
\end{center}

\begin{center}
{The MINOS Collaboration}
\end{center}


\begin{center}
            Submitted to the XXIII International Symposium \\
            on Lepton and Photon Interactions at High Energy, \\
            Daegu Korea, August 13--18, 2007.
\end{center}

\begin{abstract}
Updating our previous measurements with new data 
and analysis modifications, 
we report preliminary results on the energy-dependent deficit 
of muon-neutrinos from the Fermilab NuMI beam as observed 
with the MINOS Far Detector located $735\;$km away in the 
Soudan Underground Laboratory.   
From an exposure of $2.50\times 10^{20}$ protons on target, 
we observe 563 charged-current $\nu_\mu$ interaction 
candidates in the Far Detector,  
where $738\pm 30$ events are expected in the absence of 
neutrino oscillations.   
We have analyzed these data assuming 
two-flavor $\nu_\mu\to \nu_\tau$ oscillations.  
From a simultaneous fit to the reconstructed $\nu_\mu$ energy 
spectra obtained during two running periods 
we obtain the neutrino squared-mass difference 
$\Delta m_{32}^2 = (2.38\,\,^{+0.20}_{-0.16})\times 10^{-3}\; \mbox{eV}^2/c^4$ 
with errors at $68\%$ confidence level (CL), 
and mixing angle 
$\sin^2 (2\theta_{23}) > 0.84$ at $90\%\;$CL. \ The uncertainties 
and confidence intervals include both statistical and systematic errors.  
All results and plots presented here are {\sl preliminary}.
\end{abstract}

\section{Introduction}
\label{s-intro}

The MINOS Experiment was designed to explore the phenomenon of 
$\nu_\mu$ disappearance as observed in experiments studying 
atmospheric neutrinos~\cite{superk_etal,skle,skI} and more recently 
in the K2K accelerator-based experiment~\cite{k2k}.
The leading hypothesis for this phenomenon is neutrino oscillations, 
with $\nu_\mu \to \nu_\tau$ the likely dominant oscillation mode. 
MINOS makes use of a configurable intense neutrino source (NuMI) 
derived from $120\;\mbox{GeV}$ protons extracted from the Fermilab Main 
Injector onto a graphite target, 
and two magnetized-iron and scintillator detectors: 
a $0.98\;$kton Near Detector (ND) located on the Fermilab site approximately 
$1\;\mbox{km}$ downstream of the NuMI target and   
a $5.4\;$kton Far Detector (FD) located in the 
Soudan Underground Laboratory at a distance of $735\;\mbox{km}$. 
The NuMI beam line and 
the MINOS detectors are described in detail elsewhere~\cite{numi,minosnim}.

In Ref.~\cite{minosprl} we reported results based on data accumulated 
during the first period of NuMI operations between May 2005 and 
February 2006.  We denote this period as `Run-I'.  Far Detector data 
collected with the target in the `low-energy' (LE) beam configuration, 
corresponding 
to an exposure of $1.27\times 10^{20}$ protons on target (POT), were analyzed 
in the context of two-flavor $\nu_\mu\to\nu_\tau$ oscillations.
Oscillation parameters were obtained from a fit to the reconstructed 
charged-current (CC) $\nu_\mu$ energy spectrum: 
$\Delta m_{32}^2 = (2.74\,\,^{+0.44}_{-0.26})\times 10^{-3}\; \mbox{eV}^2/c^4$  
for the 
squared-mass difference and $\sin^2{2\theta_{23}} = 1.00_{-0.13}$ for the 
mixing angle, where only the physical region $\sin^2{2 \theta_{23}}\le 1$ was 
considered and where the uncertainties represent approximate $68\%$ 
confidence level (CL) intervals.

Following the Fermilab accelerator complex shutdown in spring 2006, NuMI
resumed operations in June 2006.  The period from then through 
July 2007 is referred to as `Run-II'.  With some modifications 
relative to that reported in Ref.~\cite{minosprl}, we have carried out 
an analysis of the LE Run-I data plus the portion of the 
LE Run-II data collected through March 2007 (denoted as 
Run-IIa), corresponding to a combined exposure of $2.50\times 10^{20}$ POT. \
In this conference contribution, we report preliminary results 
on $\nu_\mu$ disappearance from this analysis of Run-I and Run-IIa data.

\section{Summary of Analysis Steps}
\label{s-evtsel}

Most aspects of the analysis follow those described 
in Ref.~\cite{minosprl}.  Briefly, 
$\nu_\mu$ CC interactions candidates are selected from events 
in Near and Far Detector data samples with a reconstructed negatively 
charged muon. 
We employ the reconstructed $\nu_\mu$ energy ($E_\nu$) spectrum from the 
ND sample to obtain a prediction for the corresponding spectrum at the FD 
in the absence of oscillations.  The extrapolation of the ND spectrum 
to the FD accounts for the kinematic and geometrical effects 
that impart small (up to $\pm 30\%$) differences in shape between 
the two spectra.  We carry out a 
binned maximum-likelihood fit of the FD spectrum to the oscillation 
probability-weighted prediction, incorporating major systematic 
uncertainties via penalty factors.  The FD data was 
intentionally obscured during the analysis until all selection, 
fitting and systematic error estimation procedures were finalized.

The new analysis reported here, including updated results from the Run-I 
data, incorporates several improvements compared to our published 
analysis.  The most significant improvements are:
\begin{itemize}
   \item use of an upgraded neutrino interaction simulation 
          package~\cite{neugen}
         that features
         more accurate models of hadronization,  
         intranuclear rescattering and deep inelastic scattering processes.
         The combined effect of adopting the new hadronization and 
         intranuclear rescattering models is a downward shift in the 
         effective absolute energy scale of the MINOS detectors for 
         hadronic showers from neutrino interactions by amounts varying from 
         approximately $10\%$ at 2~GeV to $5\%$ at higher shower energies.

   \item a new track reconstruction algorithm, 
         which results in a $4\%$ increase in muon track reconstruction and 
         fitting efficiency. 

   \item increased acceptance achieved by including events with reconstructed 
         $E_\nu$ above $30\;\mbox{GeV}$ (corresponding to a $9.6\%$ expected 
         increase in event yield), and by expanding the fiducial volume 
         definition along the beam direction for the FD by $3.2\%$. 

   \item improved selection of $\nu_\mu$ CC events and
         rejection of neutral-current (NC) backgrounds by 
         use of a multivariate likelihood-based discriminant (PID) that 
         includes more observables than previously used.  The new 
         selection takes 
         advantage of correlations of the distributions of these 
         observables with event length.  The observables used
         are plotted in Fig.~\ref{fig:pidvars}, and distributions of 
         PID are shown in Fig.~\ref{fig:pid}.  We select events with 
         $\mbox{PID} > 0.85$.  
         The $\nu_\mu$ CC selection efficiencies and NC background 
         contamination fractions (in the null oscillation case for the FD)  
         are plotted as a function of $E_\nu$ in Fig.~\ref{fig:pideffpur}.
         Overall the efficiency for CC events has increased by 
         approximately $1\%$ 
         with respect to the cut described in Ref~\cite{minosprl}, while 
         the NC background has been reduced by more than a factor of two.
\begin{figure}[t]
  \centering\leavevmode
  \epsfxsize=2.08in  \epsfbox{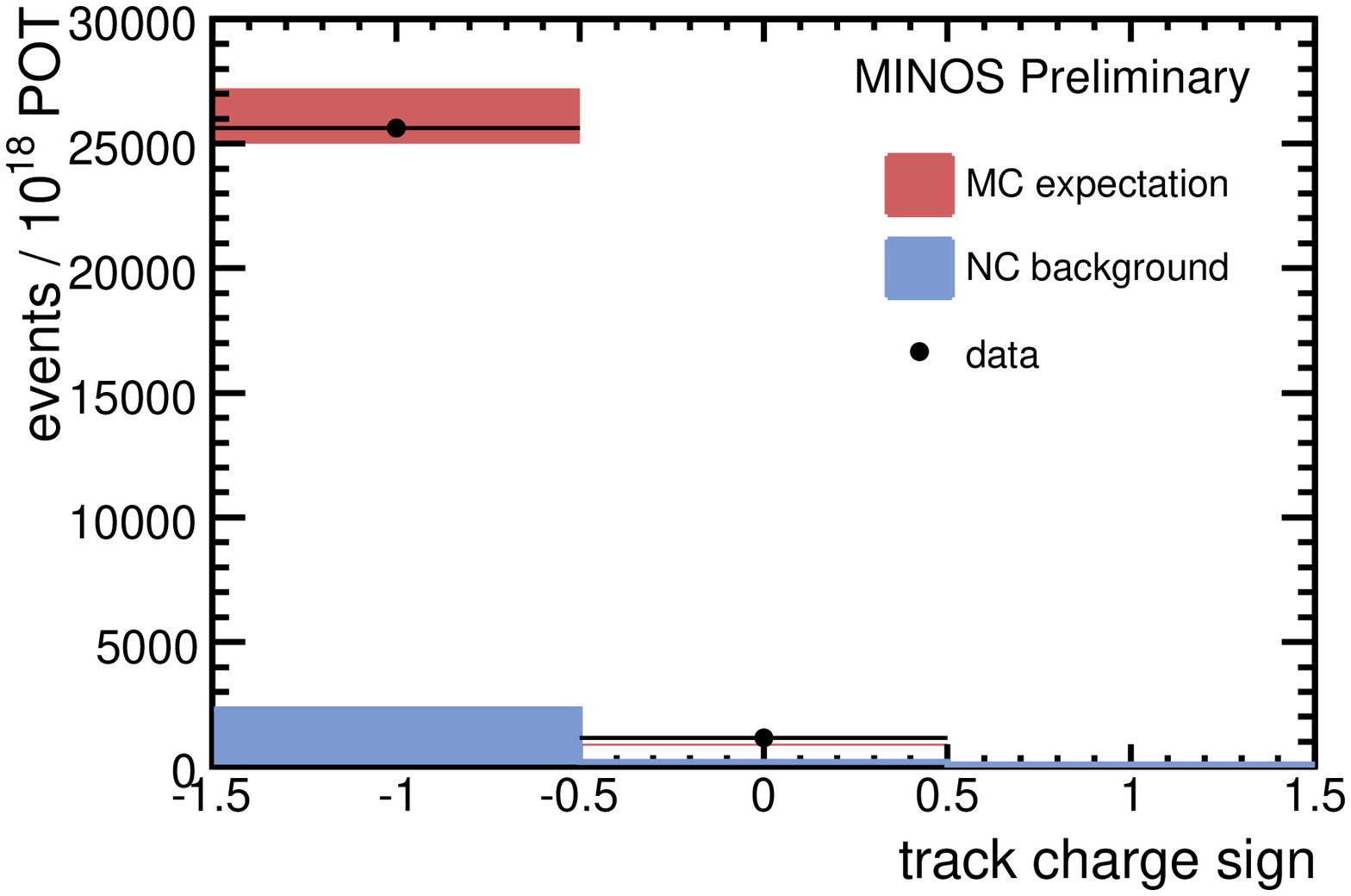}
  \epsfxsize=2.08in  \epsfbox{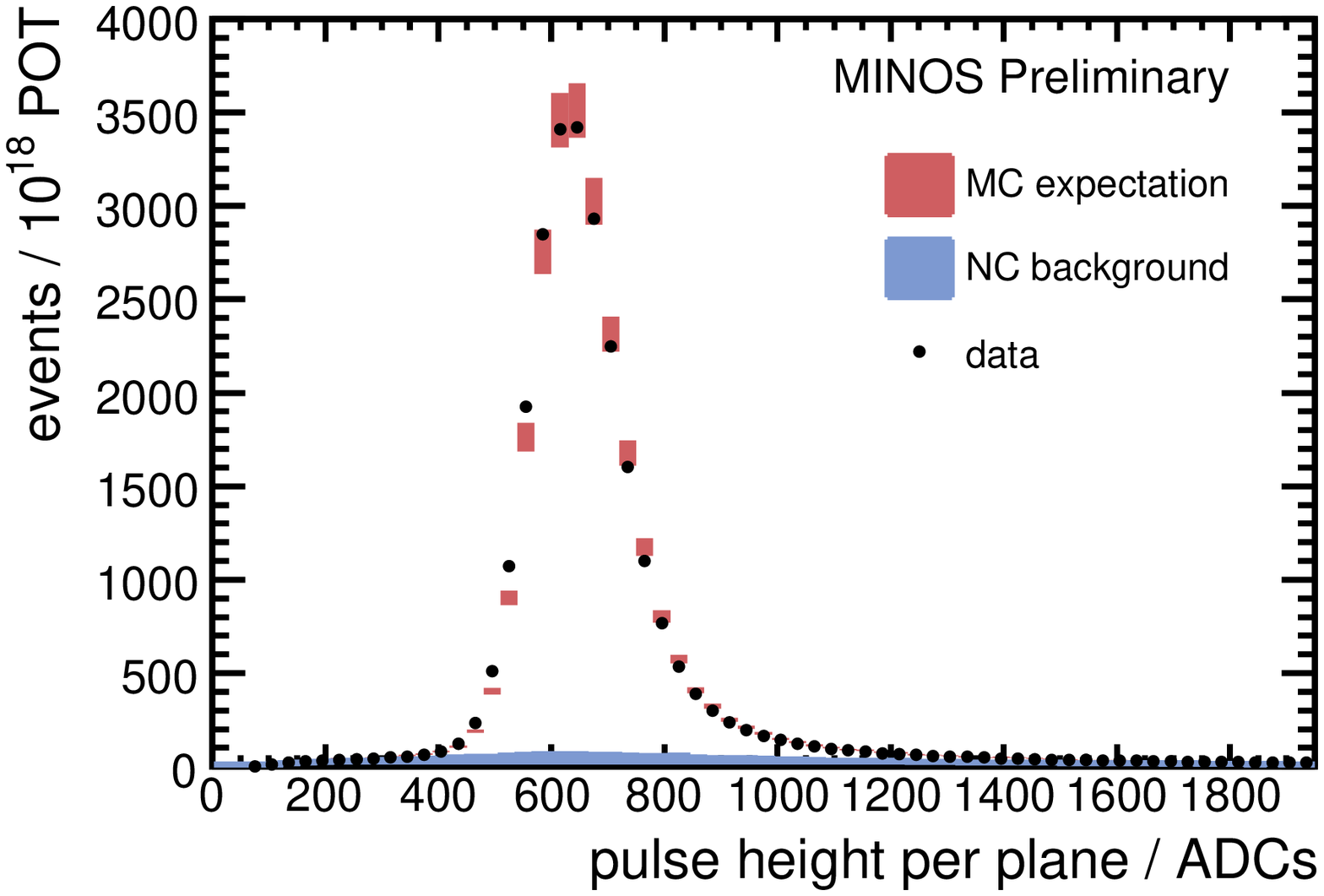}
  \epsfxsize=2.08in  \epsfbox{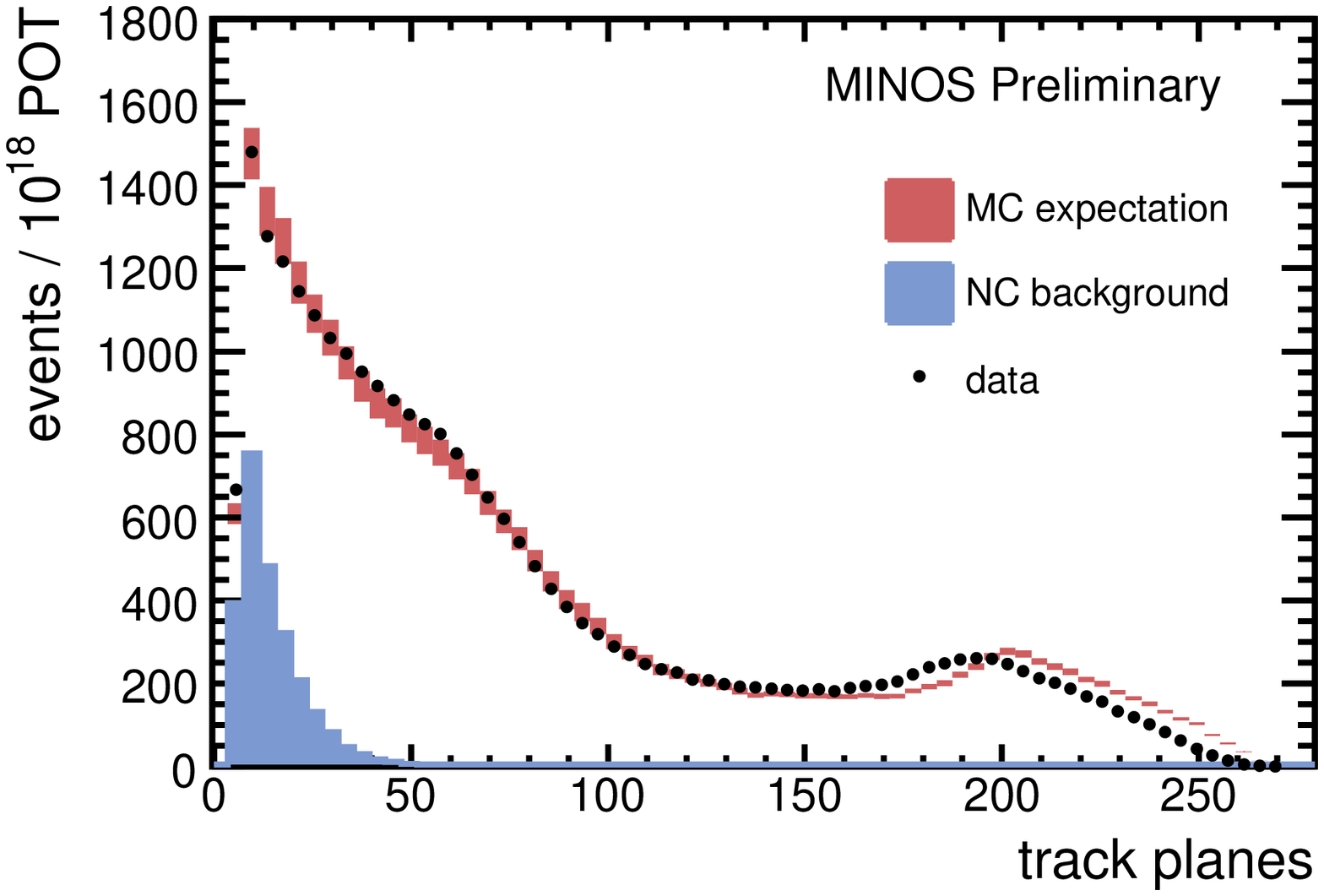}\\
  \epsfxsize=2.08in  \epsfbox{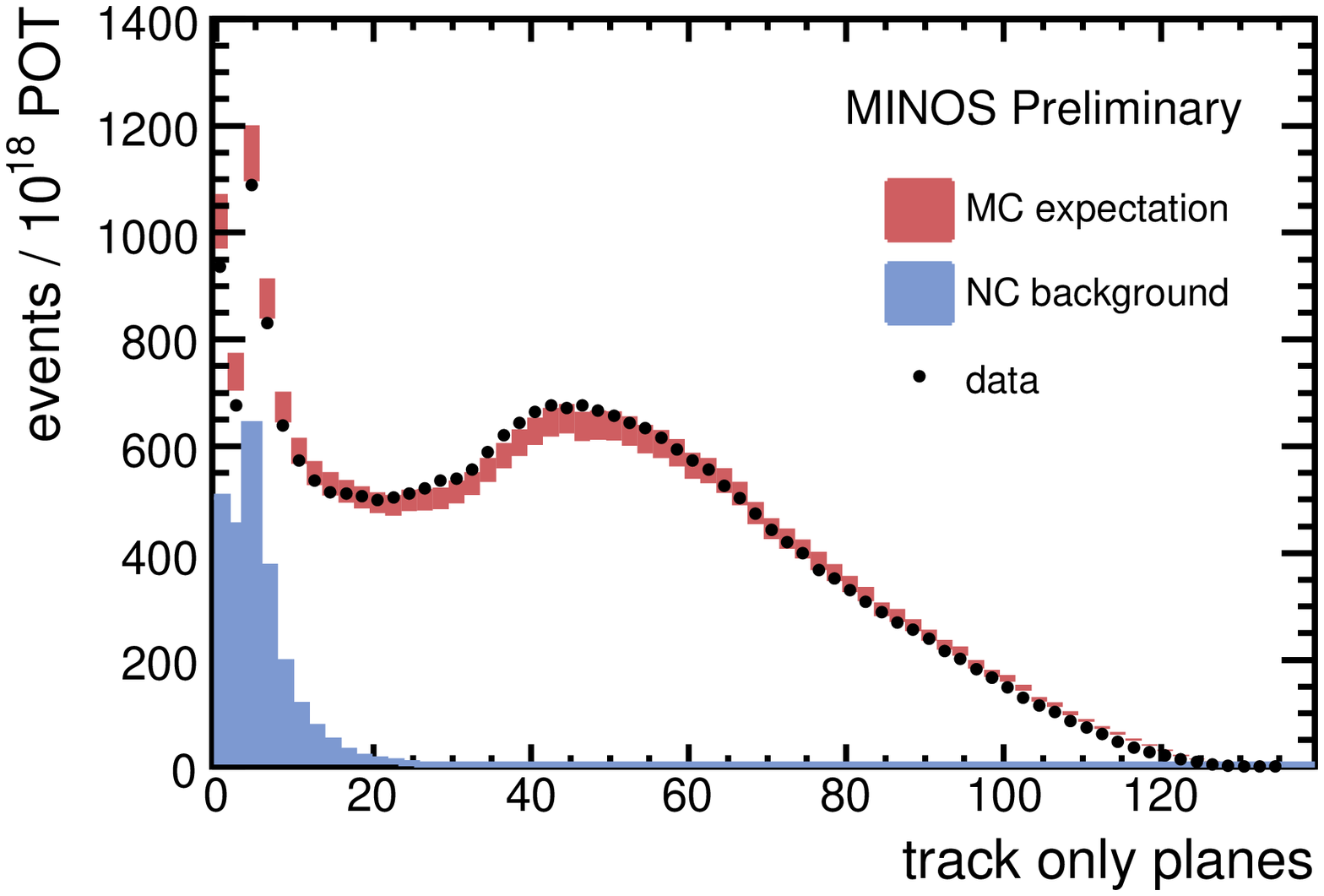}
  \epsfxsize=2.08in  \epsfbox{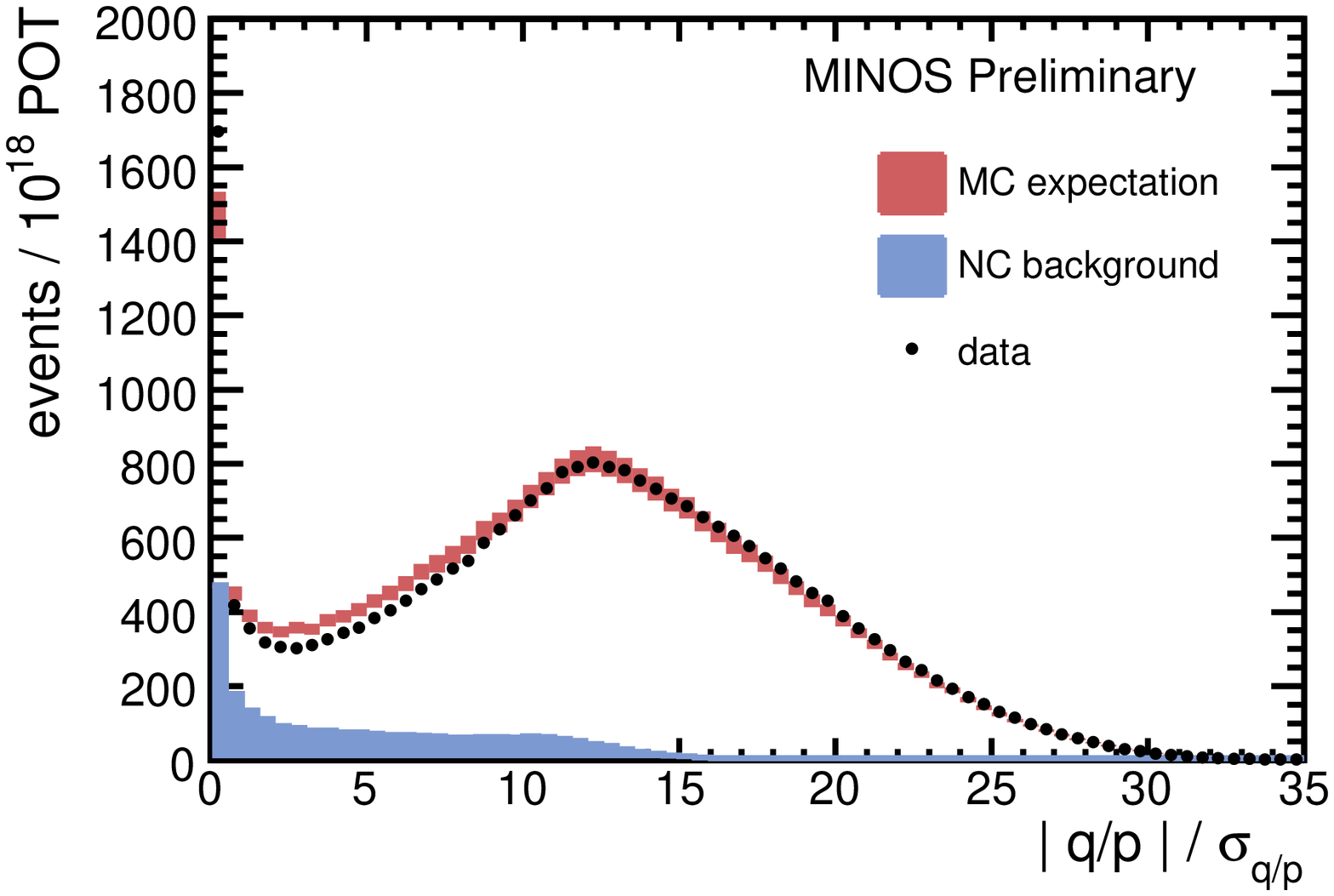}
  \epsfxsize=2.08in  \epsfbox{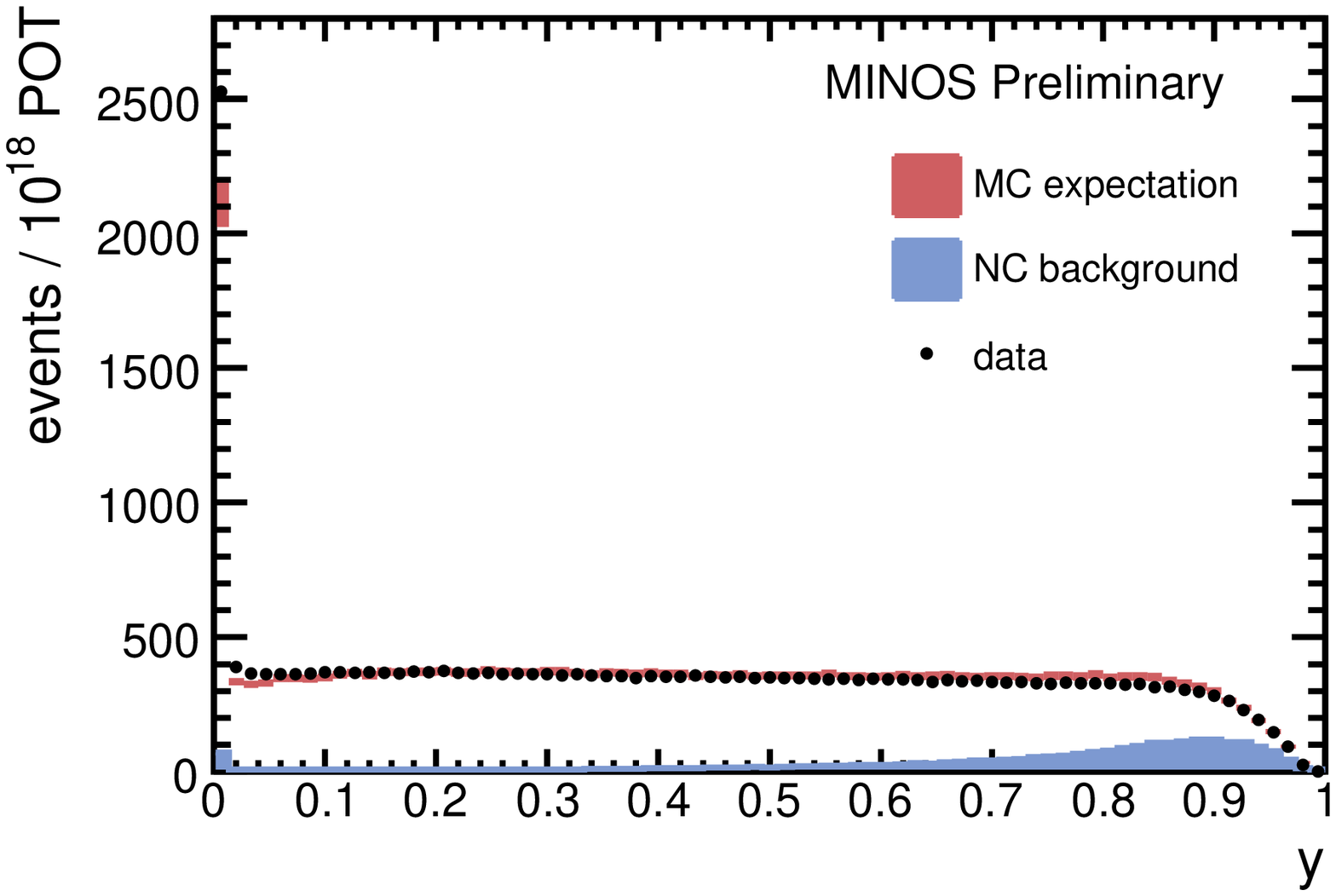}
  \caption{\small Input observables to the likelihood-ratio discriminant 
           used for CC/NC event separation are shown for ND events 
           in data (points)
           and MC (red bands).  The expected NC background contribution 
           shown by the shaded histograms.  From upper left: track charge 
           sign, average track pulse height per plane, number of 
           planes with hits on the track, number of planes with 
           hits exclusively on the track,  
           significance of track curvature measurement, 
           and reconstructed 
           $y$ defined as $E_{shower}/ E_\nu$ where $E_{shower}$ is 
           energy of the reconstructed hadronic shower.  
           In all cases the 
           MC distributions are tuned acording 
           to the results of Fig.~\ref{fig:ndbeamfit} 
           (see Sec.~\ref{s-nd}), and normalized by POT.
           The width 
           of the bands for the MC distributions reflects the 
           uncertainty associated with beam flux modeling.
}
  \label{fig:pidvars}
\end{figure}
\begin{figure}[h]
  \centering\leavevmode
  \epsfxsize=5.525in
  \epsfbox{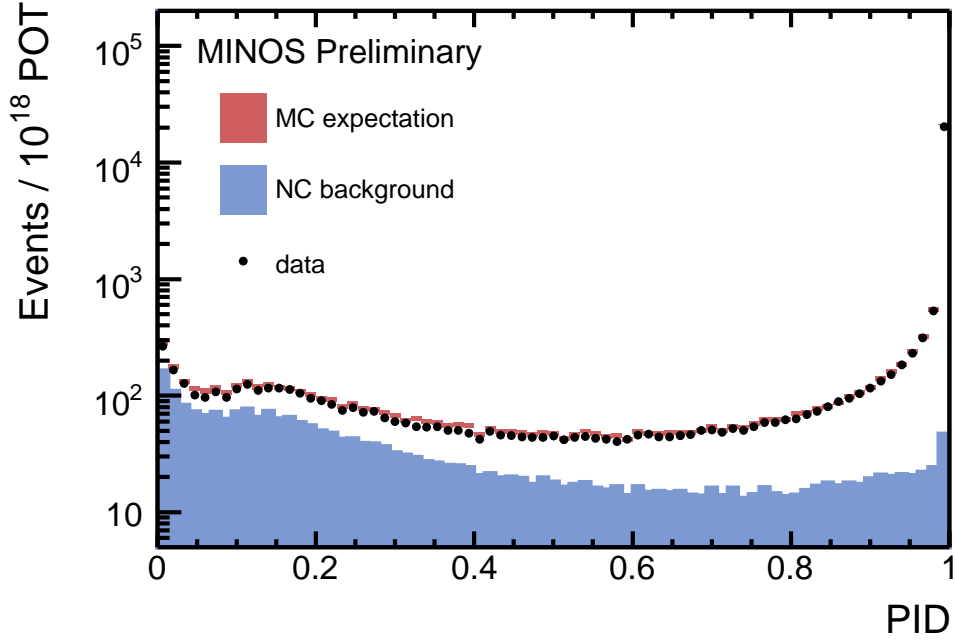}
  \caption{\small Distributions of the CC/NC separation discriminant (PID) for 
           ND data (points) and MC (red bands) samples normalized by POT 
           and weighted according to hadron production model tuning.  
           CC candidates are 
           required to satisfy $\mbox{PID} > 0.85$.  The distributions for 
           neutral current interaction events are shown represented 
           by the blue shaded historgrams.
}
  \label{fig:pid}
\end{figure}
\begin{figure}[h]
  \centering\leavevmode
  \epsfxsize=3.2in  \epsfbox{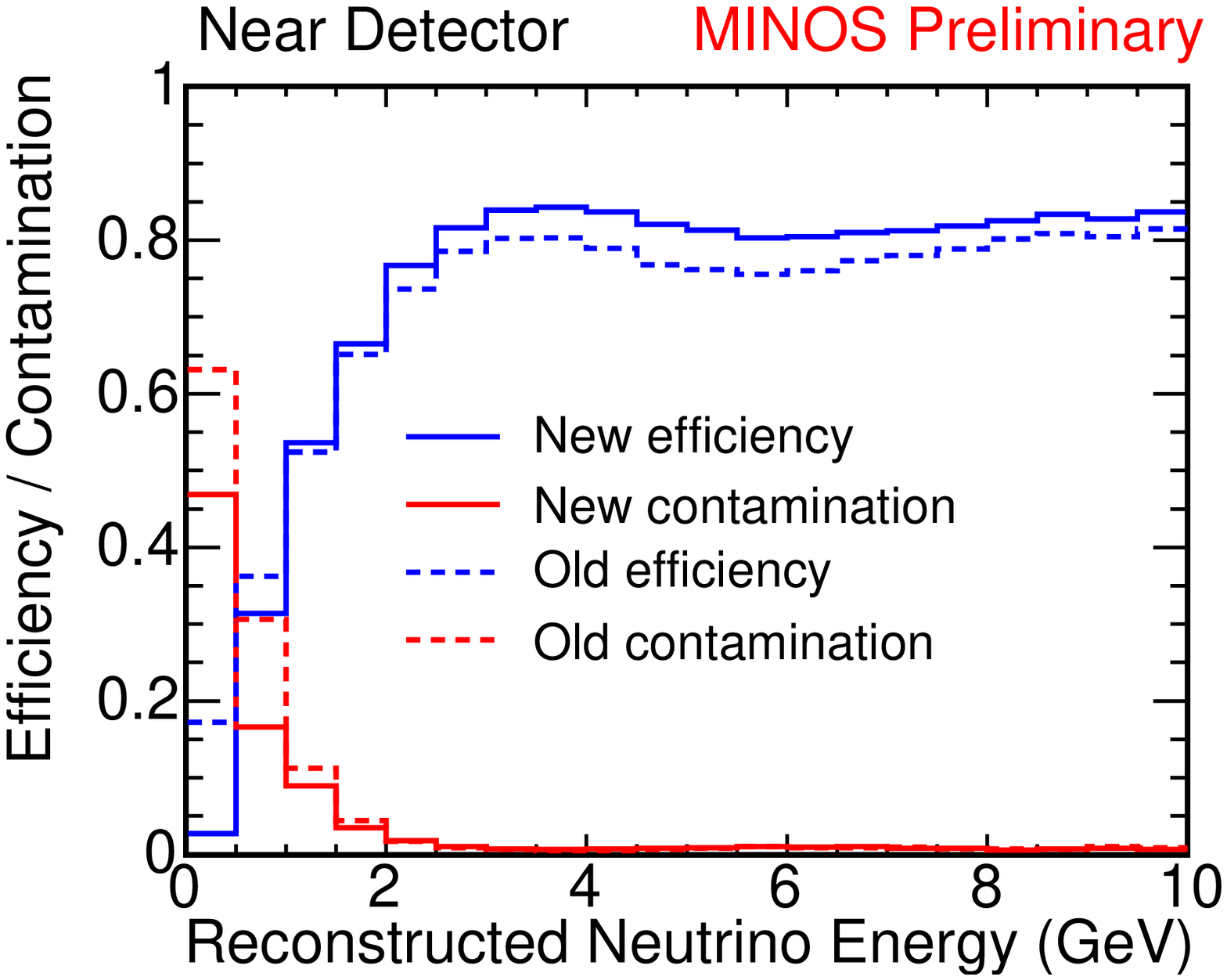}
  \epsfxsize=3.2in  \epsfbox{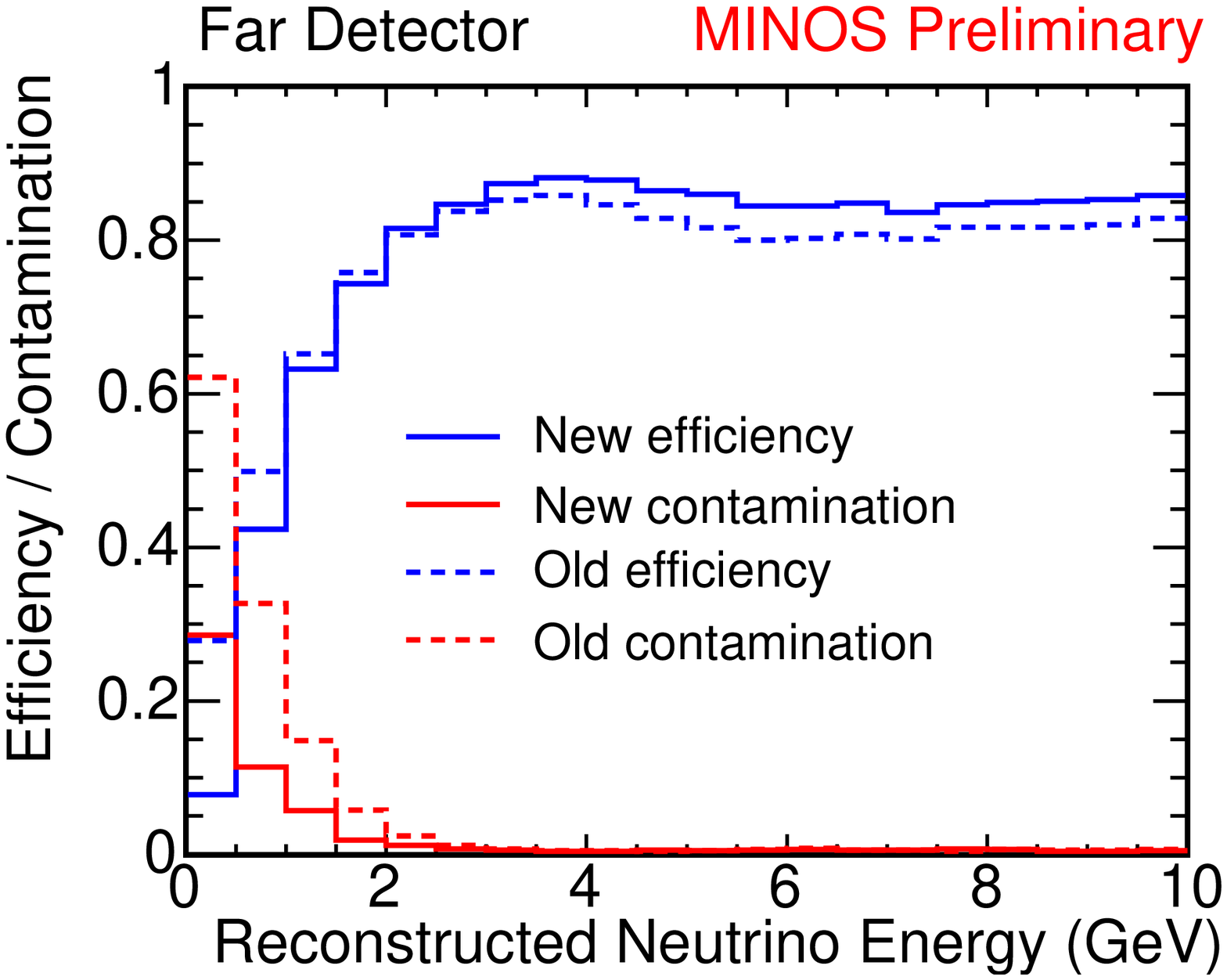}
  \caption{\small Selection efficiencies (blue) and 
           NC background contamination fractions  
           (red) 
           for the $\nu_\mu$ CC selection cut used in the new analysis 
           for the ND (left) and FD (right).  
           The corresponding curves from the selection cuts 
           used in the previous analysis~\cite{minosprl} 
           are displayed by dashed lines.  For the new analysis, 
           $\nu_\mu$ CC candidates in both ND and FD are defined by 
           $\mbox{PID} > 0.85$, whereas in the previous analysis different 
           cuts were used for the two detectors.
  }
  \label{fig:pideffpur}
\end{figure}

\end{itemize}

\clearpage
\section{\boldmath Use of ND Data to Predict the FD $E_\nu$ Spectrum}
\label{s-nd}

\begin{figure}[t]
  \centering\leavevmode
  \epsfxsize=4.875in
  \epsfbox{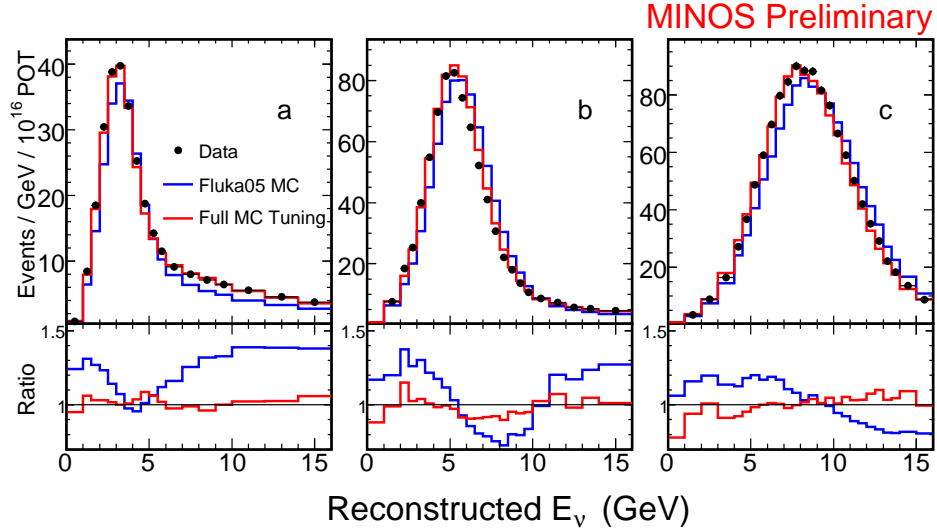}
  \caption{\small The ND reconstructed $E_\nu$ spectra (points) for 
   three of the seven beam configurations compared with 
   MC spectra obtained before (blue) and after (red)
   the hadron production and beam tuning procedure.  These 
   configurations result from varying the target location ($z$) 
   and horn current ($I$): 
   (a) low-energy beam ($z=-10\;\mbox{cm}$, relative to the nominal position 
   and $I = 185\;\mbox{kA}$), 
   (b) medium-energy beam  ($z=-100\;\mbox{cm}, \ I = 200\;\mbox{kA}$), and 
   (c) high-energy beam    ($z=-250\;\mbox{cm}, \ I = 200\;\mbox{kA}$).  
  }
  \label{fig:ndbeamfit}
\end{figure}

As in Ref.~\cite{minosprl}, we derive an expected FD $E_\nu$ spectrum 
by extrapolating the observed spectrum in the ND.  We have employed 
an extrapolation scheme, denoted the ``Beam Matrix'' method~\cite{para}, 
that is largely insensitive to mis-modeling of the neutrino flux and 
neutrino interaction cross-sections.  To correct for 
higher order effects, we have tuned the hadron production model 
and other elements of the beam flux simulation and detector response 
to improve agreement of $E_\nu$ and $E_{\overline{\nu}}$ spectra 
from data taken under seven different beam configurations with the 
corresponding MC spectra.  Fig.~\ref{fig:ndbeamfit} shows the 
$E_\nu$ spectra, comparing data with untuned 
(blue) and tuned (red) MC spectra for three beam configurations. 

\begin{figure}[ht]
  \centering\leavevmode
  \epsfxsize=6.5in
  \epsfbox{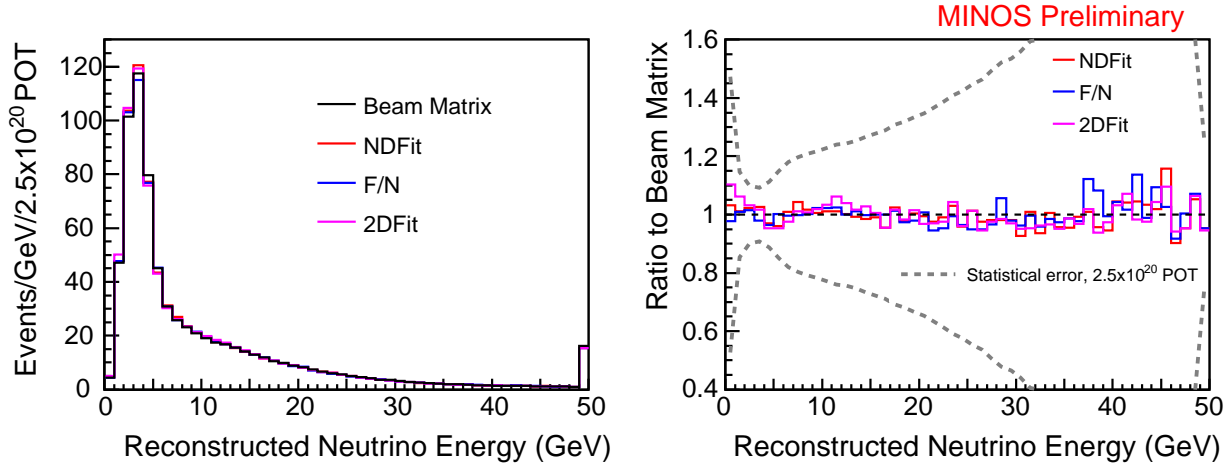}
  \caption{\small Left plot: predicted Far Detector 
  reconstructed $E_\nu$ spectra 
  from four extrapolation methods (left) based on ND data;  
  Right plot: ratios of spectra from cross-check extrapolation methods to 
  the Beam Matrix prediction.  The dashed line represents the expected 
  statistical uncertainty on FD $E_\nu$ spectrum bin contents 
  for the current exposure.  In both plots, the final bin at $49\;\mbox{GeV}$ 
  is an overflow bin, including events above $50\;\mbox{GeV}$ as well.
  }
  \label{fig:fdpred}
\end{figure}

The null-oscillation FD $E_\nu$ spectrum predicted by the Beam Matrix 
method is shown in Fig.~\ref{fig:fdpred}, along with the corresponding 
predicted spectra from three other extrapolation methods used as 
cross-checks.  Agreement between the methods is at the level of 
$4\%$.

\begin{figure}[h]
  \centering\leavevmode
  \epsfysize=2.6in
  \epsfbox{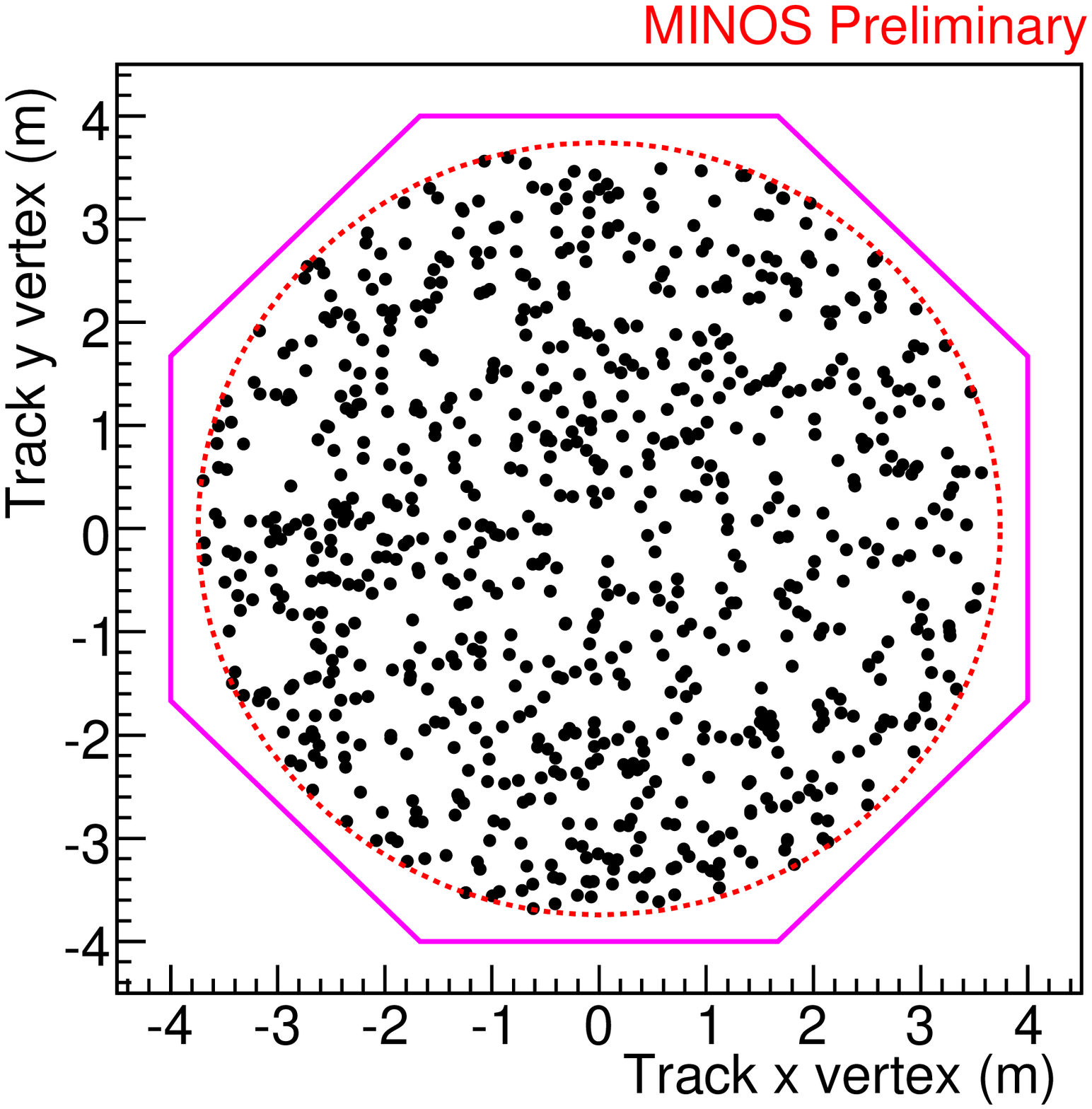}
  \epsfysize=2.6in
  \epsfbox{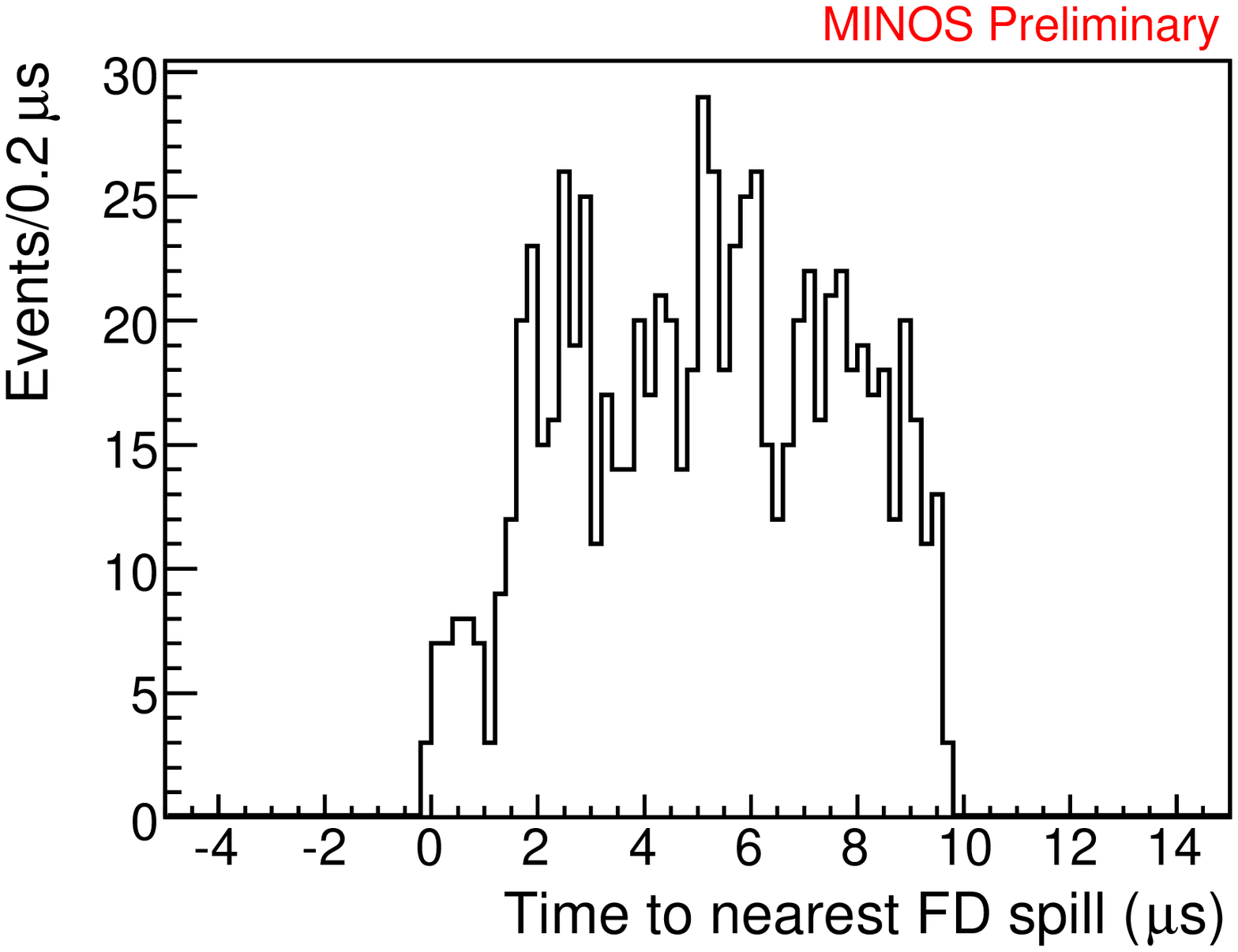}
  \caption{\small Left plot: the distribution of reconstructed 
           interaction positions 
           of neutrino-like events in the FD in the 
           transverse plane.  
           Shown also is the location of the radial fiducial volume cut.  
           Right plot: the distribution of reconstructed times of 
           FD neutrino-like
           candidates relative to the nearest NuMI beam spill time.  The 
           partial occupancy of the 0--$1.5\;\mu$s bins arises from various 
           running conditions of the Main Injector, in which the NuMI spill 
           length varied from 8.7--$10\;\mu$s, depending on other running 
           experiments.
          }
  \label{fig:fdvertex}
\end{figure}

\section{Preliminary Results}

\subsection{\boldmath FD Event Yields and $E_\nu$ Spectrum}
\label{ss-eventyields}

\begin{figure}[t]
  \centering\leavevmode
  \epsfxsize=4.225in
  \epsfbox{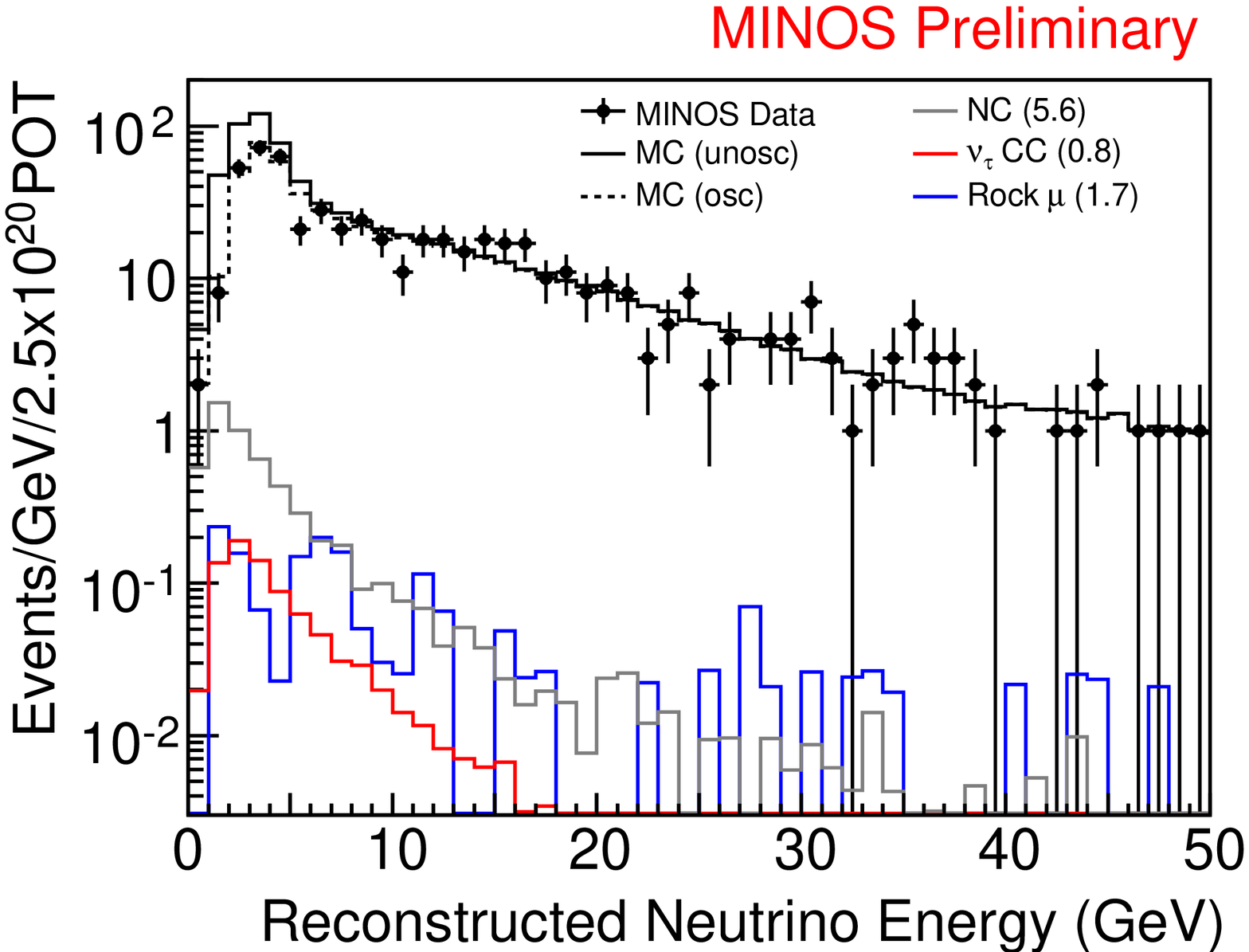}
  \caption{\small Reconstructed neutrino energy spectrum (points) from the 
  Run-I plus Run-IIa 
  Far Detector data sample. 
  Overlaid is the Monte Carlo expectation, after tuning according to 
  fits to the Near Detector data, without (solid black histogram) 
  and with (dashed) best-fit oscillation weights.  
  Shown also are the 
  expected background contributions (accounting for oscillations) 
  from $\nu_\tau$ CC events, NC events and 
  `rock muons' from beam neutrino interactions occuring in the rock 
  upstream of the 
  detector or in the detector but outside the fiducial volume.
  }
  \label{fig:fdspectra}
\end{figure}

From the Run-I and Run-IIa samples, we select 812 neutrino-like 
events with a reconstructed track in the FD fiducial volume, 
of which 563 satisfy $\nu_\mu$ CC interaction 
selection (track quality, track charge sign and NC rejection) cuts.  
Some characteristics of the neutrino-like events 
are shown in Fig.~\ref{fig:fdvertex}.
Fig.~\ref{fig:fdspectra} shows the $E_\nu$ spectrum 
for the 563 $\nu_\mu$ CC interaction candidate events (points).   
Overlaid is the null-oscillation expectation (black histogram), 
totaling $738\pm 30$ events as inferred from the ND data, 
where the uncertainty is due to the FD/ND relative normalization 
systematic error.  The expected contamination from the three main 
background sources is also shown: NC interactions (5.6 events), 
$\nu_\tau$ CC interactions (0.8 events) and $\nu_\mu$ CC interactions 
in the rock upstream of the FD or within the FD but outside the
fiducial volume (1.7 events).  The estimates for the latter 
two sources account for oscillation effects.
%

\subsection{\boldmath Oscillation Fit}
\label{s-oscanalysis}

\begin{figure}[t]
  \centering\leavevmode
  \epsfxsize=4.6in
  \epsfbox{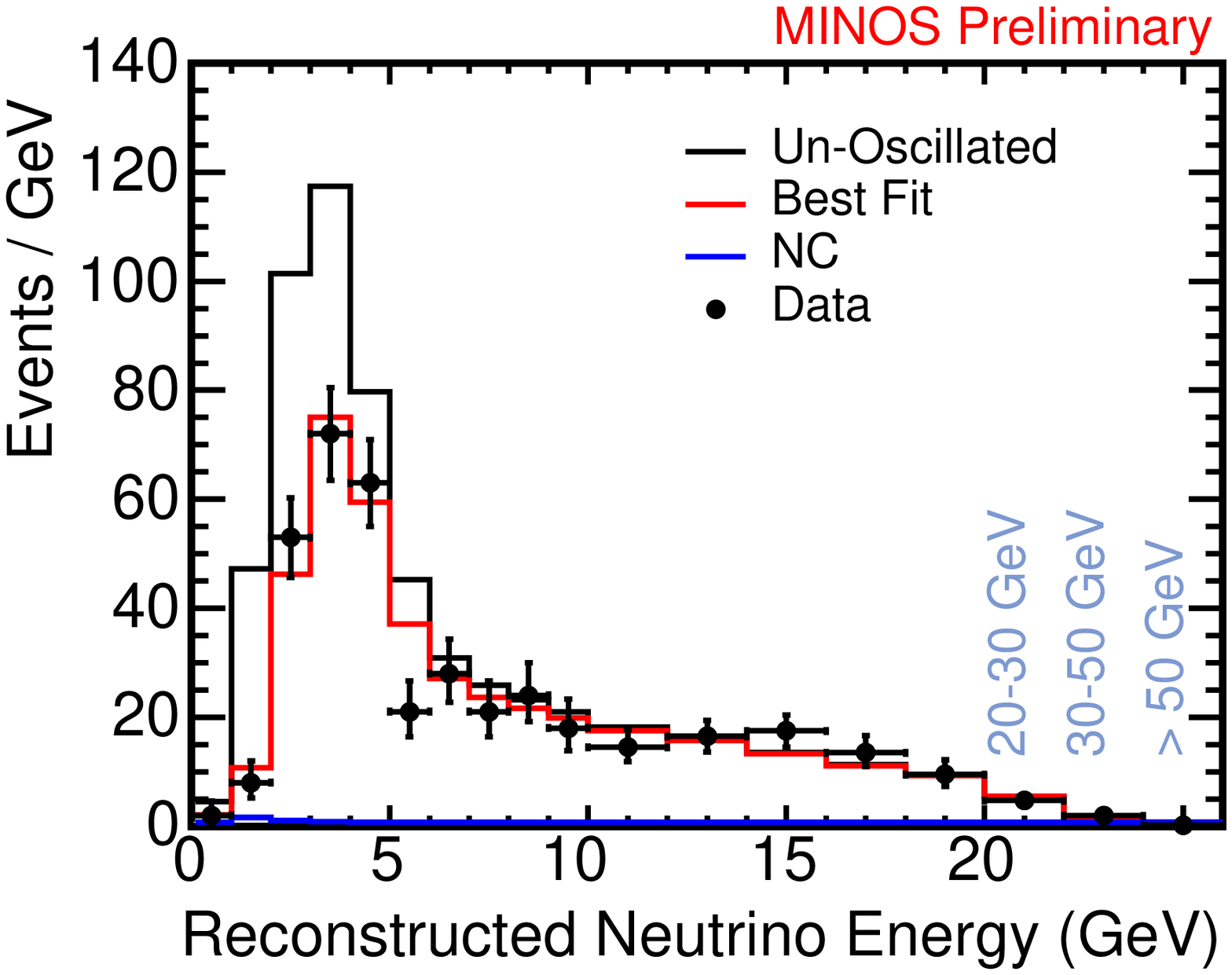}\\
  \epsfxsize=4.6in
  \epsfbox{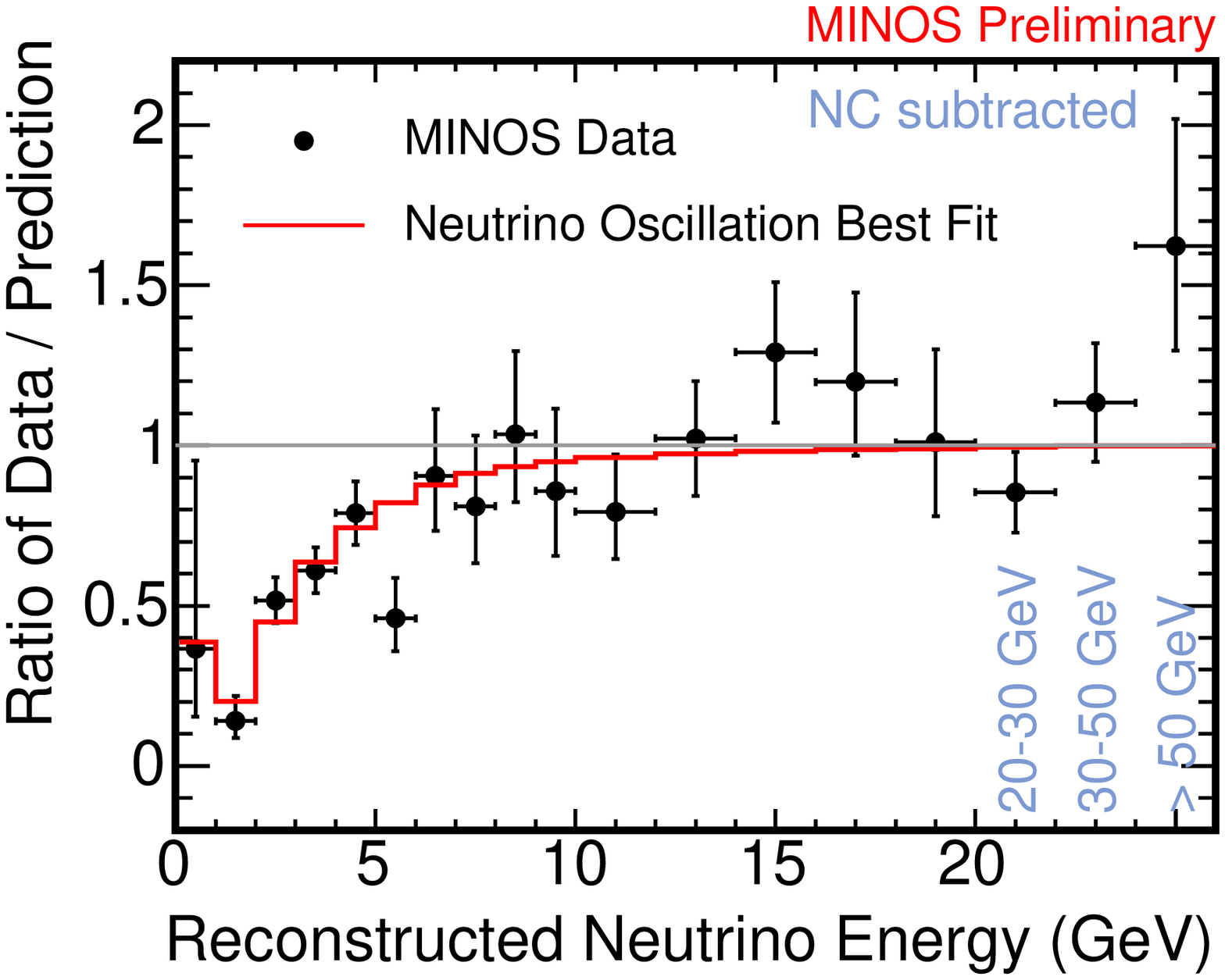}
  \caption{\small Top plot: the reconstructed $\nu_\mu$ CC energy spectrum 
  in the combined LE Run-I and IIa Far Detector data (points), with the 
  null oscillation prediction overlaid (black histogram).    
  Also shown are the oscillation-weighted prediction using 
  best-fit parameters (red) and the expected NC background contribution (blue).
  The numbers of events in the final three bins, corresponding to energy 
  ranges of 20--30, 30--50 and 50--$200\;\mbox{GeV}$, have been scaled 
  according to these bin widths.
  Bottom plot: the ratio of the NC-background subtracted FD spectrum 
  to the null-oscillation prediction (points), with the best-fit 
  oscillation expectation overlaid (red).
  }
  \label{fig:fdfits}
\end{figure}
We carried out a simultaneous fit of the oscillation-weighted 
predicted $E_\nu$ spectra for Runs I and IIa to the corresponding 
observed FD spectra.  
The separation of Run-I and IIa FD data 
is motivated by differences observed in the corresponding ND 
spectra of $\sim 7\%$ in the peak region owing to a 
difference of $1\;\mbox{cm}$ in the NuMI target placement along 
the beam axis for the two running periods.  
In the fit the quantity $\chi^2 = -2\ln{\cal L}$, where 
${\cal L}$ is the likelihood function, 
is minimized with respect to 
oscillation parameters $\Delta m^2$ and $\sin^2{2\theta}$ as well as
nuisance parameters {\boldmath $\alpha$} incorporating the 
most signficant sources of systematic uncertainty:
\[
  \chi^2(\Delta m^2, \sin^2{2\theta}, \mbox{\boldmath $\alpha$} )
      = \sum_i \left\{2(e_i -o_i) + 2o_i \,\ln(o_i/e_i)\right\} 
      + \sum_j \frac{\Delta \alpha_j^2}{\sigma^2_{\alpha_j}}\, ,  
\]
where $o_i$ represents the observed number of events in the $i^{th}$
energy bin and $e_i$ represents the corresponding oscillation-weighted 
expectation.  Only values of $\sin^2(2\theta)\le 1$ were considered. 
We included 
systematic uncertainties associated with the relative FD/ND normalization 
($\pm 4\%$), the hadronic shower energy scale ($\pm 10\%$) and 
the NC background ($\pm 50\%$).

\begin{figure}[ht]
  \centering\leavevmode
  \epsfxsize=4.8in
  \epsfbox{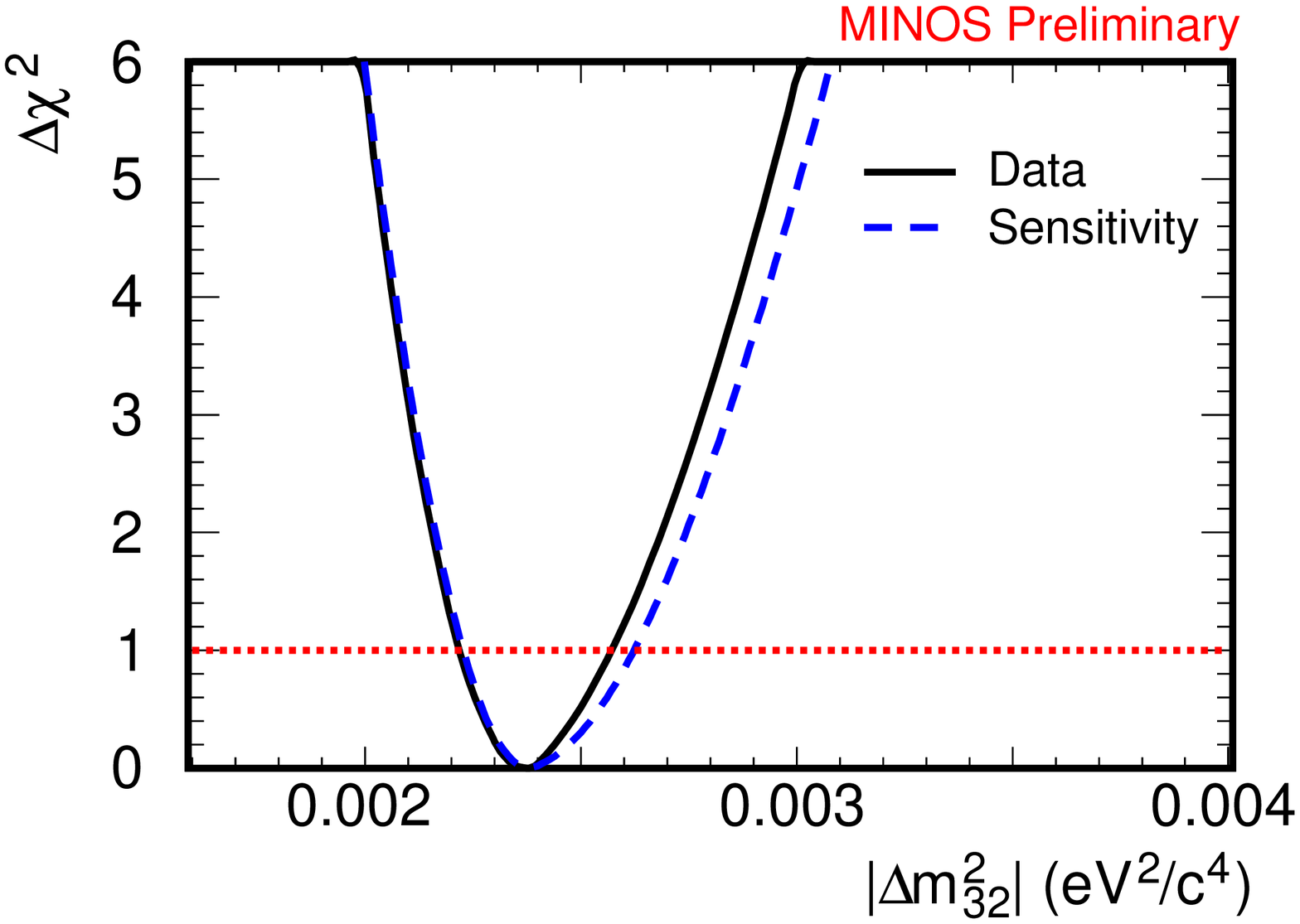}\vspace{0.15in}\\
  \epsfxsize=4.8in
  \epsfbox{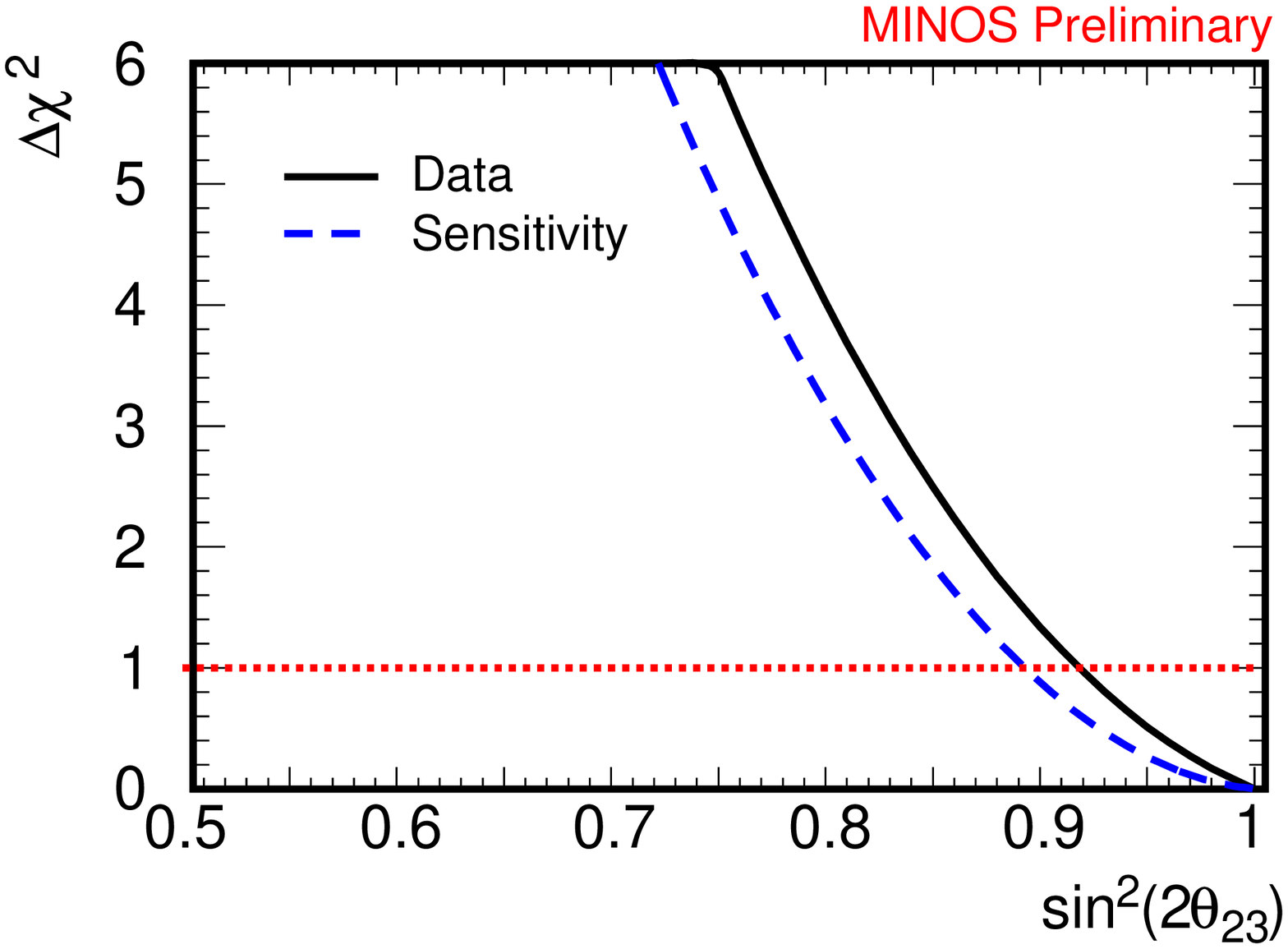}
  \caption{\small Plots of $\chi^2$ versus $\Delta m^2_{32}$ (top) and 
           $\sin^2{2\theta_{23}}$ (right) for the analysis reported 
           here (bottom).
           At each point $\chi^2$ is minimized with respect to other 
           fit parameters.  
           The corresponding sensitivity are shown as dashed blue 
           curves.
  }
  \label{fig:proj}
\end{figure}
The plots in 
Fig.~\ref{fig:fdfits} show the predicted FD spectrum weighted according 
to the best-fit oscillation parameter values (red) overlaid on the observed 
spectrum (points).  
The best-fit oscillation parameter values are: 
\begin{eqnarray*}
    \left| \Delta m_{32}^2 \right| 
    &  = & (2.38\,\,^{+0.20}_{-0.16}) \times 10^{-3}\; \mbox{eV}^2/c^4 \\
    \sin^2{2\theta_{23}} & = & 1.00_{-0.08}\, ,
\end{eqnarray*}
corresponding to $\chi^2 = 41.2$ for 34 degrees of freedom.
The uncertainties represent 68\% CL intervals, as estimated 
from the oscillation parameter value(s) giving an 
increase in $\chi^2$ of one unit relative to the best-fit value 
when minimized with respect to all other parameters.  The 90\% CL 
lower limit on $\sin^2{2\theta}$ is found to be 0.84.  

The values of $\chi^2$ relative to the best-fit value 
are plotted separately as a function of $\Delta m^2$ and 
$\sin^2{2\theta}$ in Fig.~\ref{fig:proj}.  Also shown are curves 
corresponding to the expected sensitivity, based on high-statistics 
Monte Carlo samples.  
When we relax the requirement $\sin^2{2\theta}\le 1$, the best-fit 
point moves into the unphysical region: 
$\Delta m^2 = 2.26 \times 10^{-3}\;\mbox{eV}^2/c^4$ 
and $\sin^2{2\theta} = 1.07$, with $\chi^2 = 40.9$.  This feature 
accounts for the observed trend that the obtained $\Delta \chi^2$ 
curves are more restrictive than the corresponding sensitivity curves 

\begin{figure}[t]
  \centering\leavevmode
  \epsfxsize=4.8in
  \epsfbox{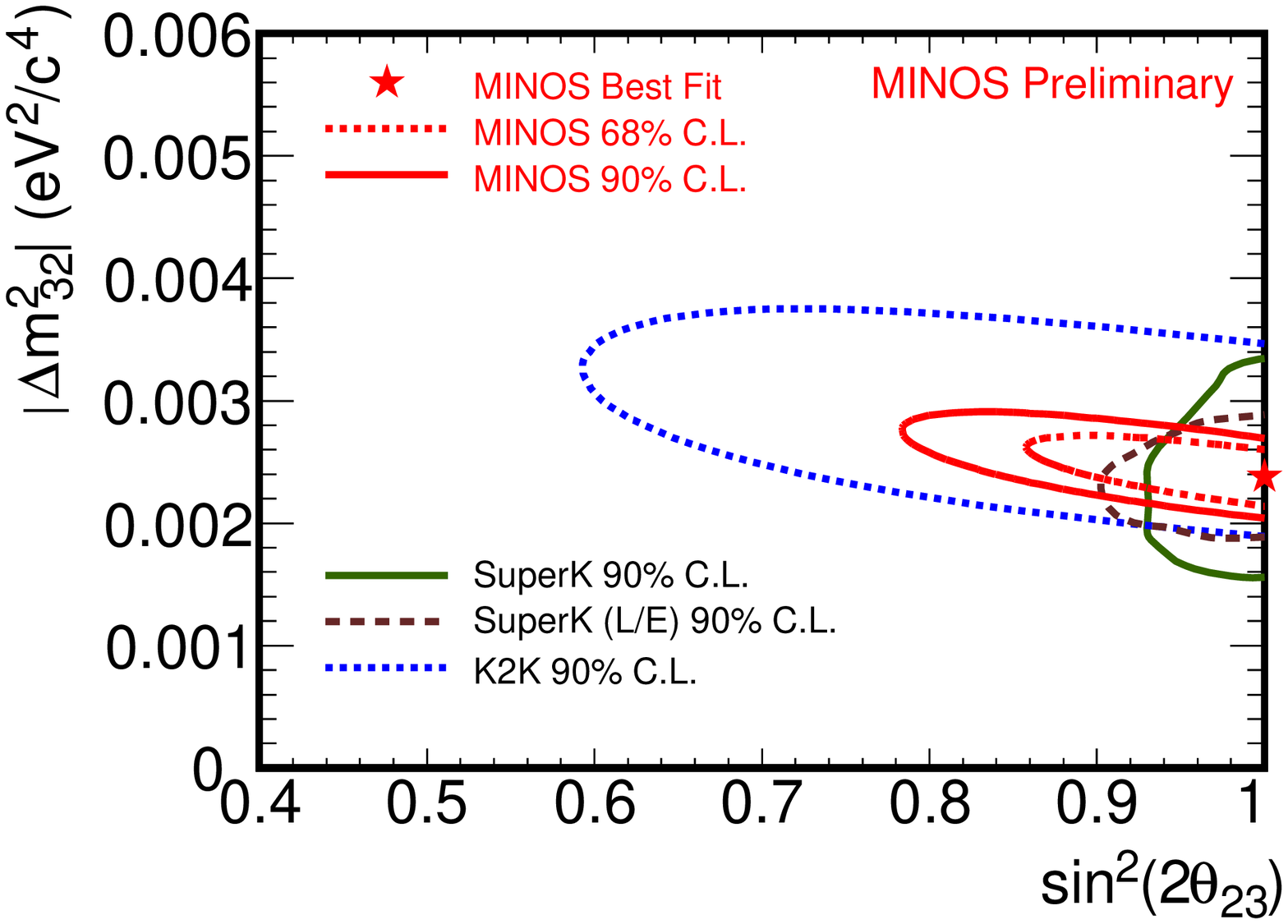}\\
  \epsfxsize=4.8in
  \epsfbox{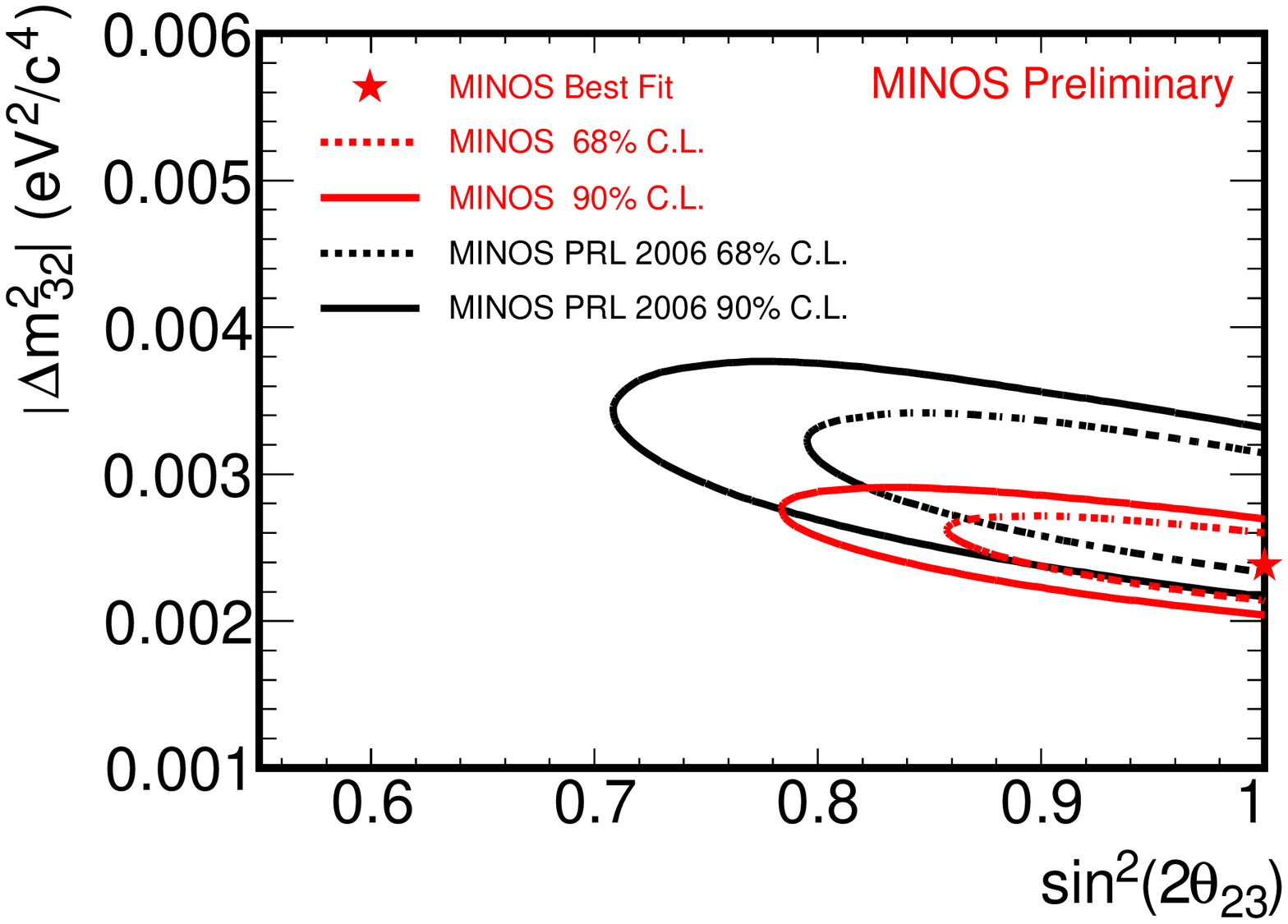}
  \caption{\small Top: The new preliminary MINOS best fit point (star) 
           and the 
           68\% and 90\% CL contours (red) as determined according to 
           $\Delta \chi^2 = 2.30$ and 4.61, respectively.  Overlaid are the 
           90\% CL contours from the Super-Kamiokande zenith 
           angle~\cite{skI} 
           and $L/E$ analyses~\cite{skle}, as well as that from the K2K 
           experiment~\cite{k2k}.  Bottom: The new preliminary MINOS contours 
           (red)
           are compared with the corresponding contours obtained in the 
           original MINOS analysis of Run-I data~\cite{minosprl}.
  }
  \label{fig:contours}
\end{figure}

The 68\% and 90\% CL contours in oscillation parameter space are shown 
in Fig.~\ref{fig:contours}.  These are specified by the locus of parameter 
values giving $\Delta \chi^2 = 2.30$ and 4.61, respectively, relative 
to the best fit point.  
We confirmed the coverage of these confidence intervals in a study 
employing the unified approach of Feldman and Cousins~\cite{ref:fc}. 

The new MINOS contours are compared with those from the published 
analysis of Run-I data~\cite{minosprl} in the right plot in 
Fig.~\ref{fig:contours}.  The allowed region is shifted to lower 
values of $\Delta m^2$ in the new analysis.  Considering the two running 
periods separately, we find  
$\Delta m^2 = (2.50\,\,^{+0.24}_{-0.20}) \times 10^{-3}\; \mbox{eV}^2/c^4$ for 
Run-I and $(2.22\,\,^{+0.44}_{-0.22}) \times 10^{-3}\; \mbox{eV}^2/c^4$ 
for Run-IIa.
The change in the absolute shower energy scale relative to the 
previous analysis (see Sec.~\ref{s-evtsel}) accounts for a systematic 
decrease of $0.06\times 10^{-3}\; \mbox{eV}^2/c^4$ for both running periods 
relative to the published Run-I analysis.  
We have also carried out the Run-I analysis with 
the new neutrino interaction and event reconstruction software, but 
applying the same selection criteria as used in the previous analysis 
so as to have a greater overlap of FD $\nu_\mu$ CC candidates, 
and obtain $\Delta m^2 = 2.46\times 10^{-3}\;\mbox{eV}^2/c^4$.  We 
estimate the statistical significance of the deviation of this result 
from our published value (after accounting for the shower energy scale 
change) to be approximately two standard deviations based on the 
MC expectations for number of FD events lost and gained in migrating 
from the old to new track reconstruction codes.

\clearpage
\section{\boldmath Summary}
\label{s-summary}

We have reported preliminary results on $\nu_\mu$ disappearance 
from the MINOS experiment based an exposure corresponding to 
$2.50\times 10^{20}$ protons on target.   
The oscillation fit results are:
\begin{eqnarray*}
    \left| \Delta m_{32}^2 \right| 
    &  =  & (2.38\,\,^{+0.20}_{-0.16}) \times 10^{-3}\; \mbox{eV}^2/c^4 \quad
            (68\%\; \mbox{CL}\; \mbox{errors})\\
    \sin^2{2\theta_{23}} & > & 0.84 \quad  (90\%\; \mbox{CL})\, , 
\end{eqnarray*}
where the uncertainty includes both statistical and systematic sources.
The value of $\Delta m^2$ is smaller than but consistent 
with the previous MINOS result~\cite{minosprl} of 
$(2.74\,\,^{+0.44}_{-0.20})\times 10^{-3}\;\mbox{eV}^2/c^4$.  
We are currently in the process of analyzing the full Run-I and Run-II 
dataset, with a total exposure of $3.5\times 10^{20}$ POT, 
of which $3.25\times 10^{20}$ POT is in the LE beam 
configuration and $0.16\times 10^{20}$ POT is in the HE configuration.

\section{\boldmath Acknowledgements}
\label{s-ack}

We thank the Fermilab staff and the technical staffs of the participating 
institutions for their vital contributions.  This work was supported by 
the U.S. Department of Energy, the U.K. Particle Physics and Astronomy 
Research Council, the U.S. National Science Foundation, the State and 
University of Minnesota, the Office of Special Accounts for Research 
Grants of the University of Athens, Greece, and FAPESP (Funda\c{c}\~ao de 
Amparo \`a Pesquisa do Estado de S\~ao Paulo) and CNPq (Conselho Nacional de 
Desenvolvimento Cientifico e Tecnologico) in Brazil.  We gratefully 
acknowledge the Minnesota Department of Natural Resources for 
their assistance and for allowing us access to the facilities 
of the Soudan Underground Mine State Park.  We also thank the 
crew of the Soudan Underground Physics laboratory for their tireless 
work in building and operating the MINOS detector.


\end{document}